\documentclass[epj]{svjour}
\usepackage{graphics}
\usepackage{epsfig}
\usepackage{amssymb}

\begin{document}
\title{Nickel isotopes in stellar matter}

\author{Jameel-Un Nabi\inst{1,2}}
\institute{
  \inst{1} Faculty of Engineering
Sciences, GIK Institute of Engineering Sciences and Technology,
Topi 23640, Swabi, Khyber Pakhtunkhwa, Pakistan\\
  \inst{2} The Abdus Salam ICTP,
Strada Costiera 11, 34014, Trieste, Italy }

\abstract{Isotopes of nickel play a key role during the silicon
burning phase up to the presupernova phase of massive stars.
Electron capture rates on these nickel isotopes are also important
during the phase of core contraction. I present here the microscopic
calculation of ground and excited states Gamow-Teller (GT) strength
distributions for key nickel isotopes. The calculation is performed
within the frame-work of pn-QRPA model. A judicious choice of model
parameters, specially of the Gamow-Teller strength parameters and
the deformation parameter, resulted in a much improved calculation
of GT strength functions. The excited state GT distributions are
much different from the corresponding ground-state distributions
resulting in a failure of the Brink's hypothesis. The electron
capture and positron decay rates on nickel isotopes are also
calculated within the framework of pn-QRPA model relevant to the
presupernova evolution of massive stars. The electron capture rates
on odd-A isotopes of nickel are shown to have dominant contributions
from parent excited states during as early as silicon burning
phases. Comparison is being made with the large scale shell model
calculation. During the silicon burning phases of massive stars the
electron capture rates on $^{57,59}$Ni are around an order of
magnitude bigger than shell model rates and can bear consequences
for core-collapse simulators.
}

\authorrunning{Nabi}
\titlerunning{Nickel isotopes in stellar matter}
\maketitle

\onecolumn
\section{Introduction}
Collective excitation properties of nuclei at finite temperature is
of great utility in astrophysical environments. Of special mention
is the behavior of charge-exchange transitions in hot nuclei. The
Gamow-Teller (GT) transition is one of the most important nuclear
weak processes of the spin-isospin ($\sigma \tau$) type. Ikeda,
Fujii and Fujita \cite{Ike63} predicted the GT resonance in 1962
which was later discovered by the Indiana group using the (p,n)
reaction \cite{Gaa80}. Since then the interest in the study of
spin-isospin symmetry and the GT strength properties of the nuclei
increased manyfold (see for example the review article by Ref.
\cite{Ost92}).

The GT properties of nuclei in the medium mass region are main
prerequisites to study the presupernova evolution of massive stars
\cite{Bet90}. Throughout this paper massive stars imply stars with
mass $M \ge 10 M_{\odot}$. As soon as the core of a massive star
exceeds the appropriate Chandrasekhar mass, electrons are captured
by nuclei in stellar matter and the electron capture rate is a
crucial factor that determines the initial collapse phase. The
GT$_{+}$ transitions contribute significantly in the determination
of electron capture rates during the presupernova phase of massive
stars. In the GT$_{+}$ strength, a proton is changed into a neutron
(the plus sign is for the isospin raising operator ($t_{+}$),
present in the GT matrix elements, which converts a proton into a
neutron). In many collapse simulations the GT$_{+}$ strength was
assumed to reside in a single resonance and the strength of this
resonance was determined from the single-particle model
\cite{Ful80}. The approximation was done mainly due to insufficient
experimental information and rather limited theoretical and
computational advancements prevailing at the time. Since then
considerable progress was made, both in the theoretical and
experimental side, which led to a better understanding of the
nuclear structure. Aufderheide and collaborators \cite{Auf93} showed
that a strong phase space dependence makes the relevant electron
capture rates more sensitive to GT distributions than to total GT
strengths. The (n,p) experiments on the other side revealed the fact
that the GT$_{+}$ strength is fragmented over many states, and that
the total strength is quenched compared to single-particle model.
Microscopic calculations of GT$_{+}$ strength functions and
weak-interaction rates in stellar matter were then performed
successfully using the pn-QRPA model (e.g. \cite{Nab99}) and the
shell model (e.g. \cite{Lan00}).

Isotopes of nickel play inarguably a crucial role in the
presupernova evolution of massive stars. Many simulation studies of
presupernova evolution of massive stars showed electron capture
rates on nickel isotopes to considerably alter the lepton-to-baryon
rate of change of the stellar core (e.g. \cite{Auf94,Heg01}).
Aufderheide and collaborators \cite{Auf94} also showed the
$\beta$-decay on few neutron-rich nickel isotopes to affect the
lepton-to-baryon ratio of stellar core. However these $\beta$-decays
were orders of magnitude smaller than electron capture rates and
were not considered important for presupernova evolution of massive
stars by Heger and collaborators \cite{Heg01}. Electron capture
rates on $^{56,57,58,59,60,61,63,64,65}$Ni were considered important
from previous simulation studies of core-collapse. The GT$_{+}$
strength distribution and electron capture rates on $^{56}$Ni were
discussed in detail earlier \cite{Nab05}. In this paper I would like
to report on the improved calculation of the GT$_{+}$ strength
functions and associated electron capture rates on nickel isotopes
(mass number 57 to 65) using the pn-QRPA as the base model. The
paper is written as follows. Section~2 briefly discusses the pn-QRPA
model and presents the calculations of GT$_{+}$ strength functions
and electron capture rates on these nickel isotopes.  The results
are compared with the large scale shell model results in Section~3.
Conclusions are given in Section~4.

\section{The nuclear model and results}
The pn-QRPA model was first developed by Halbleib and Sorensen
\cite{Hal67}. The extension of the original model to deformed nuclei
was first given by Krumlinde and M\"{o}ller \cite{Kru84} and to
excited state contributions by Muto and collaborators \cite{Mut92}.
The model was then adapted to calculate stellar weak rates (see for
example \cite{Nab99}). Two important parameters of the pn-QRPA model
are the GT strength parameters and the deformation parameter.
Special emphasis was given in the current project to smartly choose
these key model parameters. The Hamiltonian of the model was taken
to be of the form:
\begin{equation}
H^{QRPA} =H^{sp} +V^{pair} +V_{GT}^{ph} +V_{GT}^{pp}.
\end{equation}
The single particle energies and wave functions were calculated in
the Nilsson model (which takes into account nuclear deformations).
The BCS model was used to calculate the pairing force. In the
pn-QRPA formalism, proton-neutron residual interactions occur as
particle-particle (pp) and particle-hole (ph) interaction. The pp
interaction is of paramount importance for electron capture and
positron decay rate calculation \cite{Sta90,Hir93}. Both pp and ph
interaction terms are given a separable form in the model and are
characterized by two interaction constants ($\chi$ for the ph force
and $\kappa$ for the pp force).  The calculation of $\beta$-decay
half-lives depends heavily on the choice of these interaction
parameters \cite{Sta90,Hir93}. In order to improve the calculation a
fine-tuning of these strength parameters was performed for the
nickel isotopes (mass number 50 to 94). Experimental values are
available in literature for the centroids and total GT strength for
the even-even isotopes of nickel, $^{58,60,62,64}$Ni. The idea was
to find the optimal value of $\chi$ and $\kappa$ that best
reproduced the measured values for these even-even isotopes of
nickel. For $^{58}$Ni experimental data was taken from
\cite{Rap83,Wil95,Hag04,Col06,Sas09}, for $^{60}$Ni from
\cite{Rap83,Wil95,Ana08}, for $^{62}$Ni from \cite{Wil95,Ana08} and
finally for the case of $^{64}$Ni measured values were taken from
\cite{Wil95,Pop07,Pop09}. The optimum value for $\chi$ and $\kappa$
was chosen to be 0.001 MeV and 0.052 MeV, respectively. These
parameters were then fixed for the calculation of GT strength
functions and associated weak rates for all nickel isotopes. As
mentioned earlier deformation of nuclei was taken into account in
the current calculation. The deformation parameter is yet another
key pn-QRPA model parameter (see also \cite{Ste04}). For the case of
even-even isotopes, the experimentally adopted value of the
deformation parameters, extracted by relating the measured energy of
the first $2^{+}$ excited state with the quadrupole deformation, was
taken from Raman et al. \cite{Ram87}. For the case of odd-A nickel
isotopes the deformation of the nucleus was calculated as
\begin{equation}
\delta = \frac{125(Q_{2})}{1.44 (Z) (A)^{2/3}},
\end{equation}
where $Z$ and $A$ are the atomic and mass numbers, respectively and
$Q_{2}$ is the electric quadrupole moment taken from Ref.
\cite{Moe81}. Q-values were taken from the mass compilation of Audi
et al. \cite{Aud03}.

Once the model parameters were carefully chosen, the pn-QRPA model
was used for the calculation of GT strength distributions (both
ground and excited states), electron capture and positron decay
rates in stellar matter. The use of a separable interaction assisted
in the incorporation of a luxurious model space of 6 major
oscillator shells. The basic formalism for the calculation of
electron capture and positron decay rates in the pn-QRPA model can
be seen in detail from Ref. \cite{Nab04}. Around 300 parent excited
states and as many daughter excited states were considered in the
current calculation covering an energy range of up to 15-20 MeV.
Each state was treated as a resonance state having a finite
band-width of 0.044 MeV for the even-even isotopes and 0.054 MeV for
the odd-A isotopes. The GT strength distribution was calculated for
all these 300 states in a microscopic fashion.  The total electron
capture (ec) and positron decay (pd) rate per unit time per nucleus
was calculated as
\begin{equation}
\lambda^{ec(pd)} =\sum _{ij}P_{i} \lambda _{ij}^{ec(pd)},
\end{equation}
where $P_{i}$ is the probability of occupation of parent excited
states and follows the normal Boltzmann distribution and $\lambda
_{ij}$ are the partial rates. After the calculation of all partial
rates for the transition $i \rightarrow j$ the summation was carried
out over all initial and final states until satisfactory convergence
was achieved in the rate calculation. An average quenching factor of
0.6 was adopted in the current calculation.

Table~\ref{ta1} shows the value of the deformation parameter for the
nickel isotopes used in the current calculation. Shown also are the
total calculated GT strength in both (electron capture and
$\beta$-decay) directions, $\Sigma S_{\beta^{\pm}}$. The sum is
taken up to 12 MeV in daughter nuclei. It can be seen from
Table~\ref{ta1} that Ikeda sum rule was satisfied in the calculation
(small difference arise due to the fact that cutoff energy is taken
to be 12 MeV here and also due to some rounding errors).

Since the calculations of stellar electron capture and positron
decay rates depend primarily on the calculation of B(GT$_{+}$)
strength distributions, I present here the comparison of measured
B(GT$_{+}$) strength distributions of nickel isotopes
($^{58,60,62,64}$Ni) with the calculated distributions. The measured
values of B(GT$_{+}$) strength, till 8 MeV in daughter, for
$^{58}$Ni, $^{60}$Ni, $^{62}$Ni and $^{64}$Ni, is 3.8 $\pm$ 0.4,
3.11 $\pm$ 0.08, 2.53 $\pm$ 0.07 and 1.72 $\pm$ 0.09,  respectively
\cite{Wil95}. This is to be compared with the pn-QRPA calculated
values of 6.95, 5.83, 3.58 and 1.77, respectively. Authors in
\cite{Ana08} also calculated the experimentally determined value of
B(GT$_{+}$) strength for $^{60}$Ni ($^{62}$Ni), up to 2.4 (2.3)MeV
in $^{60}$Co ($^{62}$Co) to be 1.34 $\pm$ 0.22 (1.28 $\pm$ 0.29).
This is to be compared with the pn-QRPA calculated value of 2.89 and
1.63, respectively.

Table~\ref{ta2} shows the values of pn-QRPA calculated centroids for
the B(GT$_{+}$) strength distributions of nickel isotopes. For
comparison the values of centroids calculated by LSSM \cite{Lan00}
and those by Pruet and Fuller \cite{Pru03} are also shown (where
available). The values in column 3 and 4 were adapted from Ref.
\cite{Pru03}. Experimental centroids for the even-even isotopes of
nickel are given in the last column. With the exception of
$^{62}$Ni, the pn-QRPA calculated centroids are in reasonable
comparison with the measured values. The pn-QRPA model does not
exactly reproduce the measured data. However the fair comparison of
the calculated centroids and total strength functions (which control
the calculated electron capture and positron decay rates) with the
measured data does indicate that the nuclear model employed in this
calculation should represent a fair estimate of weak rates in
stellar matter where no experimental data is available and where the
overall physics is rather poorly understood. It is hoped that this
calculation could prove beneficial for core-collapse simulators
world-side.

\begin{figure}[htbp]
\caption{Gamow-Teller strength distributions in $^{57,58,59}$Ni. The
abscissa represents energy in cobalt isotopes.} \label{fig1}
\begin{center}
\begin{tabular}{c}
\includegraphics[width=0.6\textwidth]{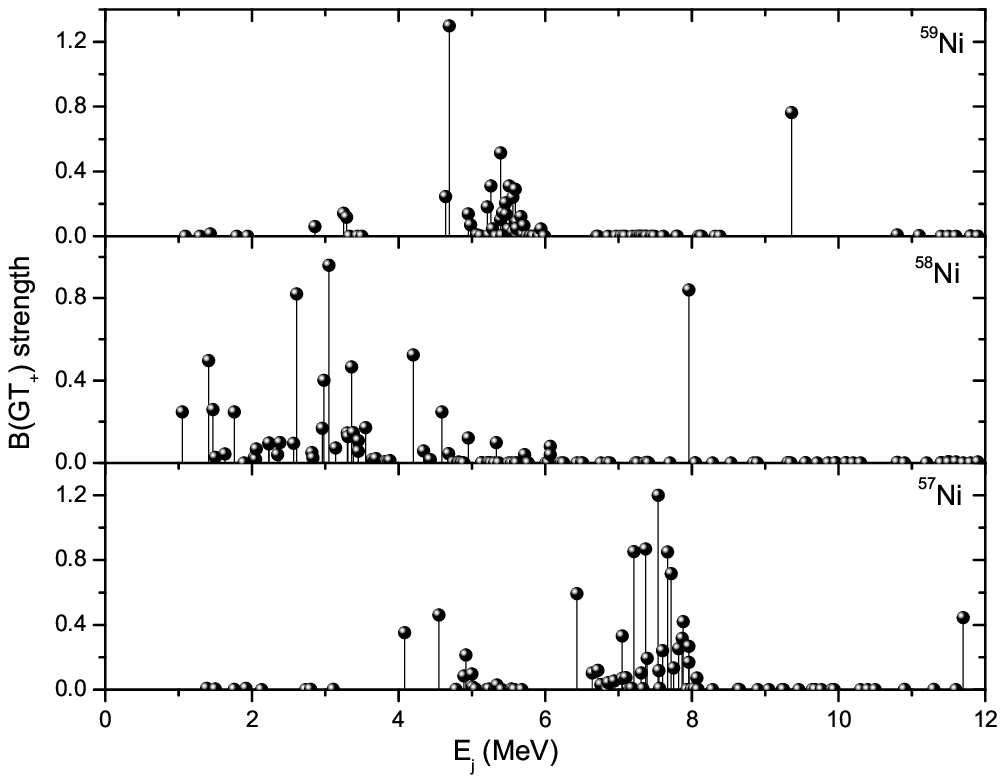}
\end{tabular}
\end{center}
\end{figure}

\begin{figure}[htbp]
\caption{Gamow-Teller strength distributions in $^{60,61,62}$Ni. The
abscissa represents energy in cobalt isotopes.} \label{fig2}
\begin{center}
\begin{tabular}{c}
\includegraphics[width=0.6\textwidth]{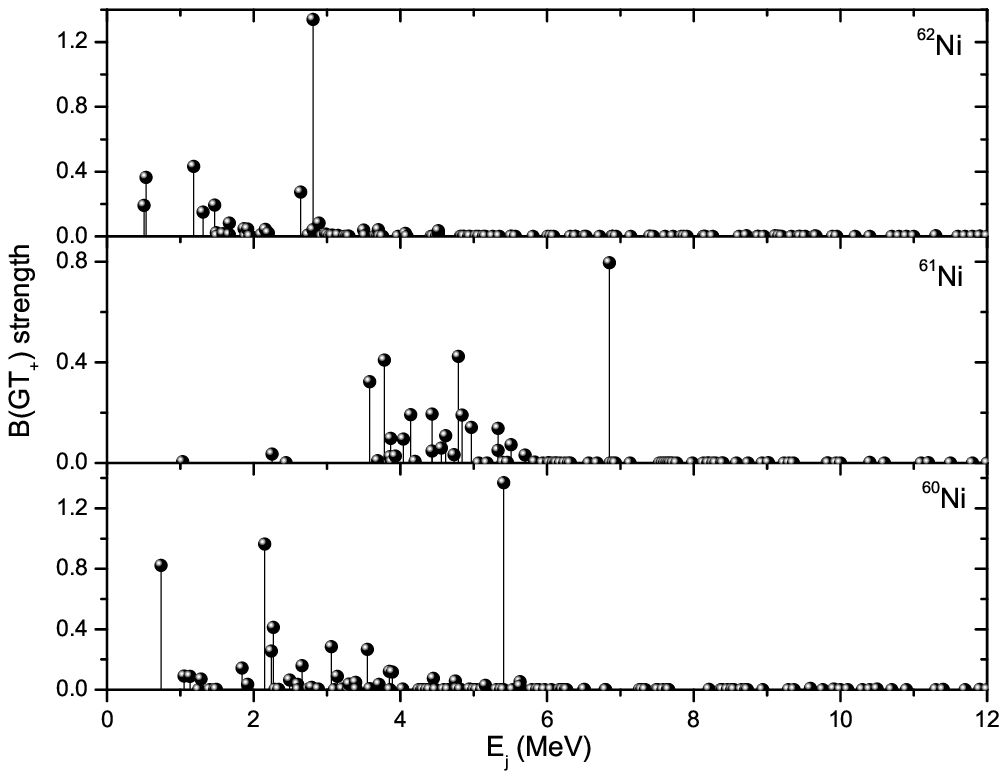}
\end{tabular}
\end{center}
\end{figure}

\begin{figure}[htbp]
\caption{Gamow-Teller strength distributions in $^{63,64,65}$Ni. The
abscissa represents energy in cobalt isotopes.} \label{fig3}
\begin{center}
\begin{tabular}{c}
\includegraphics[width=0.6\textwidth]{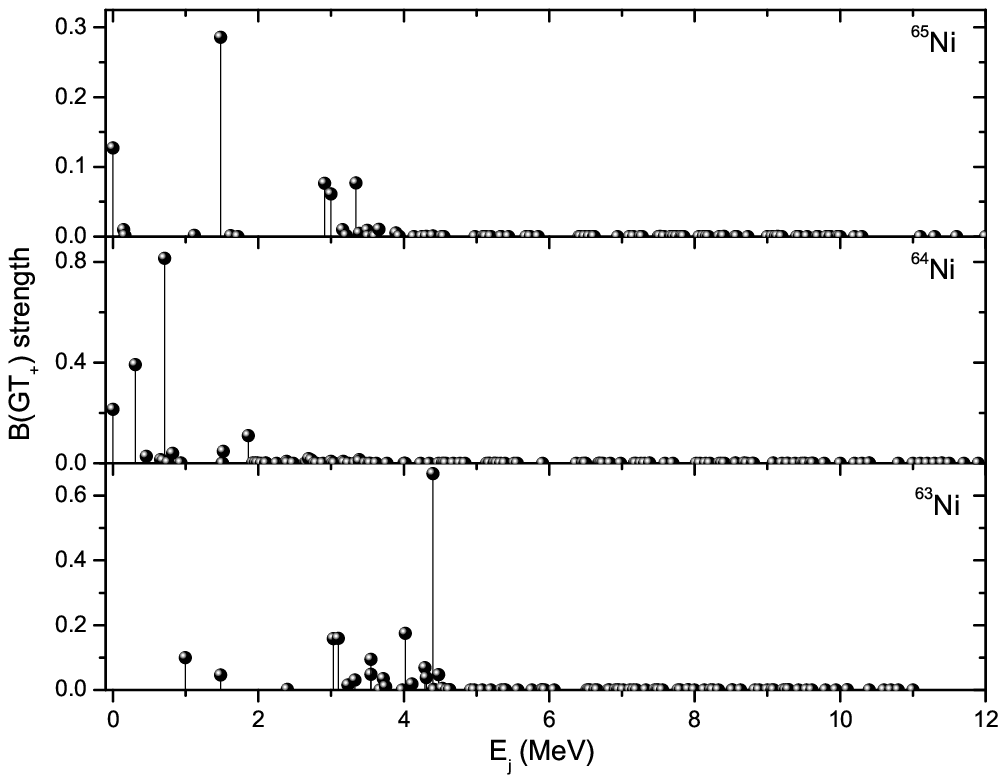}
\end{tabular}
\end{center}
\end{figure}

Figures~1--3 show the ground state B(GT$_{+}$) strength
distributions for the isotopes of nickel. Figure~\ref{fig1} shows
the calculated ground state B(GT$_{+}$) strength functions for
$^{57,58,59}$Ni. The abscissa represents energy in daughter
$^{57,58,59}$Co. Experimental data were incorporated in the
calculation wherever possible. The calculated excitation energies
were replaced with measured levels when they were within 0.5 MeV of
each other. Missing measured states were inserted and inverse
transitions (along with their log$ft$ values) were also taken into
account. No theoretical levels were replaced with the experimental
ones beyond the excitation energy for which experimental
compilations had no definite spin and/or parity. No forbidden
transitions were calculated in this project. Work is currently
underway to calculate these transitions. It can be seen from
Figure~\ref{fig1} that the model calculates many high-lying
transitions in $^{57}$Co (calculated centroid is around 7.25 MeV).
The model calculates low-lying GT transitions in $^{58}$Co (except
for the peak at 7.96 MeV) and locates the centroid at 3.57 MeV. Two
big transitions at 4.69 MeV and 9.36 MeV in $^{59}$Co moves the
B(GT$_{+}$) centroid at 5.63 MeV for the case of $^{59}$Ni.
Figure~\ref{fig2} shows the calculated strength functions for
$^{60,61,62}$Ni. Prominent peaks are seen at 5.41 MeV, 6.85 MeV and
2.81 MeV in daughter $^{60}$Co, $^{61}$Co and $^{62}$Co,
respectively. The calculated GT strength functions for
$^{63,64,65}$Ni are shown in Figure~\ref{fig3}. Here much of the
strength lies in the low-energy range in daughter nuclei. It can be
seen from Figures~1--3 that the GT strength in daughter nuclei is
well fragmented. The excited state B(GT$_{+}$) strength functions
were also calculated in a microscopic fashion and are not shown here
due to space limitations. These are seen to be much different from
the ground state distributions and imply that the Brink's hypothesis
is not a good approximation to use in calculation of stellar weak
rates for nickel isotopes (Brink's hypothesis states that GT
strength distribution on excited states is \textit{identical} to
that from ground state, shifted \textit{only} by the excitation
energy of the state). It would be shown later that electron captures
on nickel isotopes have significant contributions from these excited
states during the presupernova and core contraction phases of
massive stars. Further electron capture rate on odd-A isotopes of
nickel, $^{57,59,61,63}$Ni, have dominant contributions from parent
excited states during as early as silicon burning phases. The ASCII
files for the B(GT$_{\pm}$) strength distributions for ground and
all excited states of nickel isotopes can be requested from the
author.

\begin{figure}[htbp]
\caption{(Color online) Electron capture and positron decay rates on
$^{57,58,59,60}$Ni as a function of stellar temperature and density.
All rates are given in units of $s^{-1}$. Densities are given in
units of $gcm^{-3}$ and temperatures in units of $10^{9}$ K.}
\label{fig4}
\begin{center}
\begin{tabular}{c}
\includegraphics[width=0.6\textwidth]{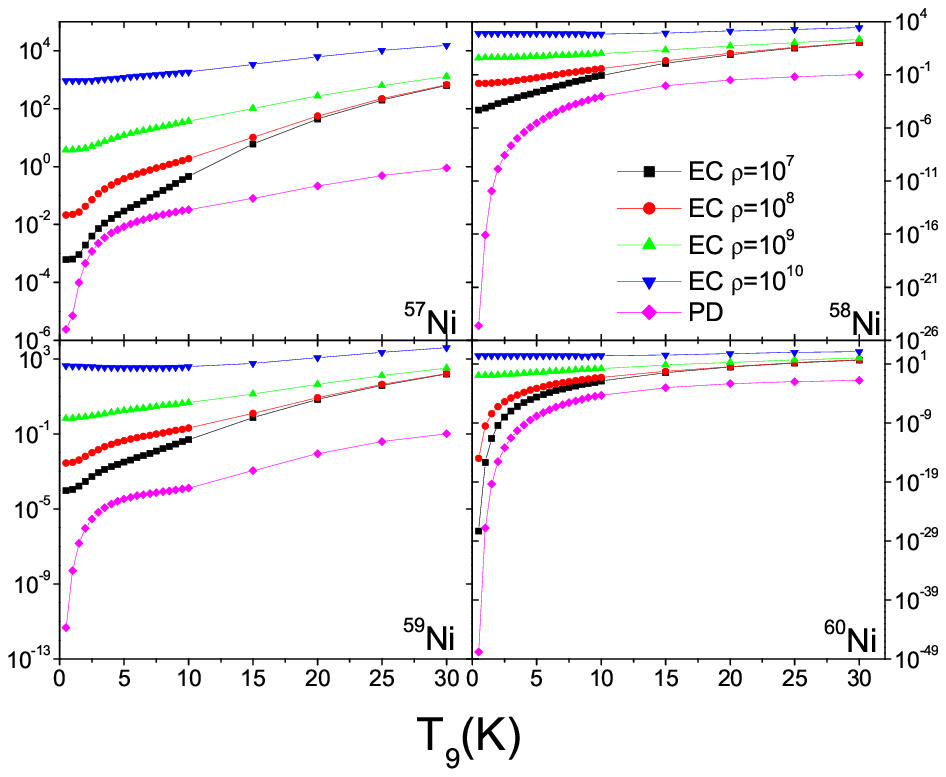}
\end{tabular}
\end{center}
\end{figure}

\begin{figure}[htbp]
\caption{(Color online) Electron capture and positron decay rates on
$^{61,63,64,65}$Ni as a function of stellar temperature and density.
All rates are given in units of $s^{-1}$. Densities are given in
units of $gcm^{-3}$ and temperatures in units of $10^{9}$ K.}
\label{fig5}
\begin{center}
\begin{tabular}{c}
\includegraphics[width=0.6\textwidth]{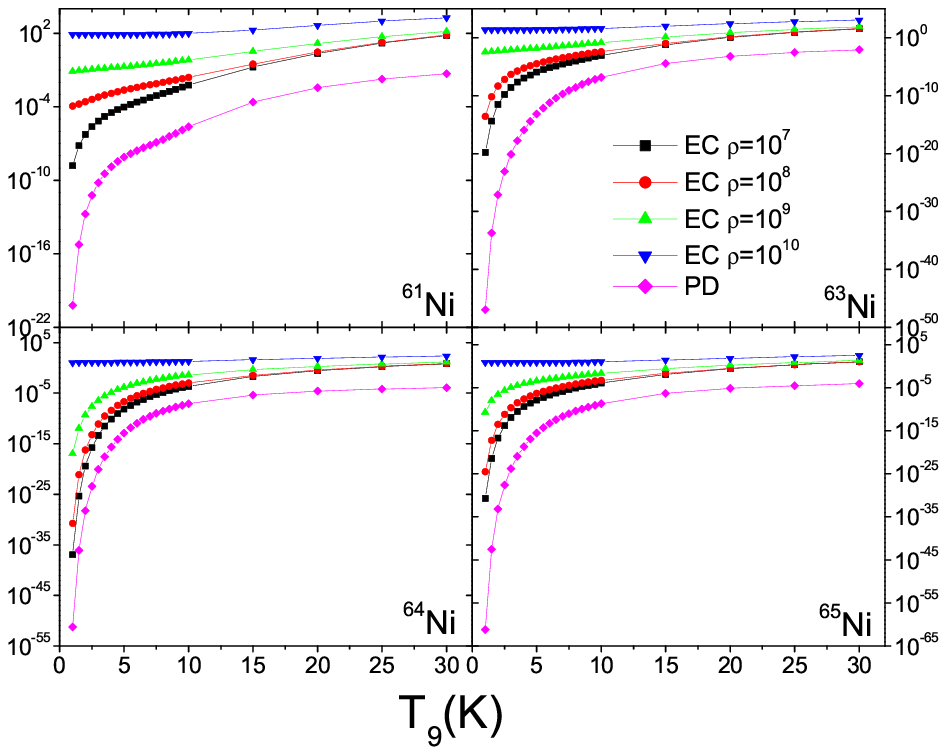}
\end{tabular}
\end{center}
\end{figure}

The pn-QRPA calculated electron capture and positron decay rates of
nickel isotopes are shown in Figures~4--5. Figure~\ref{fig4} depicts
the electron capture and positron decay rates of $^{57,58,59,60}$Ni
as a function of stellar temperature and density. The electron
capture rates are shown for density scale $\rho = 10^{7,8,9,10}$
gcm$^{-3}$. T$_{9}$ gives the stellar temperature in units of
$10^{9}$ K. The temperature and density scale chosen are pertinent
to silicon burning phase to core contraction phase of massive stars.
The calculated positron decay rates are independent of stellar
density and much smaller than competing electron capture rates. It
can be seen from Figure~\ref{fig4} that electron capture rate
increases both with increasing stellar density and with increasing
stellar temperature. The centroid for the GT$_{+}$ strength
distribution shifts to lower excitation energy in daughter nucleus
with increasing nuclear temperature. As density increases the
fraction of electrons with energy sufficient for excitation of the
GT$_{+}$ resonance increases in the electron gas. The positron decay
rates on even-even isotopes of nickel, $^{58,60}$Ni are smaller than
$^{57,59}$Ni by many orders of magnitude. Except for the case of
$^{57}$Ni (during the silicon burning phases), the positron decay
rates can be safely neglected  in comparison to the electron capture
rates by simulators. Figure~\ref{fig5} shows corresponding
weak-interaction rates for the case of $^{61,63,64,65}$Ni. Electron
capture rates on $^{62}$Ni are not considered important from
previous simulation results and are as such not presented here
(interested readers can request these rates from the author). For
the isotopes of nickel, $^{61,63,64,65}$Ni, once again it is seen
that the positron decay rates are much smaller and may be neglected
in simulation codes compared with the competing electron capture
rates. The electron capture rates increase with increasing
temperature and density for reasons mentioned above. The complete
set of electron capture and beta decay rates on a detailed
temperature-density grid point (suitable for interpolation purposes
and simulation codes) for nickel isotopes can be requested from the
author.

In order to analyze the contribution of the parent excited states to
the total rates, the ratio of ground state capture rates to total
capture rates, $R_{ec}(G/T)$, was calculated for the nickel
isotopes. Table~\ref{ta3} shows this contribution at a selected
density scale of $\rho = 10^{8.5}$ gcm$^{-3}$ as a function of
increasing stellar temperature. Similar results were obtained for
lower densities (pertaining to silicon burning phases) and higher
densities (pertaining to presupernova and core contraction phases).
It can be seen from Table~\ref{ta3} that for the odd-A nickel
isotopes, $^{57,59,61,63}$Ni, excited states have dominant
contribution at T$_{9} \ge 3 $K.

To explain further I take the sample case of $^{57}$Ni. Here the
first excited state is located at 0.77 MeV. For the ground state the
calculated values of the GT strength in electron capture direction,
$\Sigma S_{\beta^{+}}$, and the centroid is 9.98 and 7.25 MeV,
respectively (up to 12 MeV in daughter). For the first excited state
the calculated corresponding values are 10.9 and 6.73 MeV,
respectively. (A similar comparison of pn-QRPA calculated ground and
excited states GT strength distributions for iron isotopes were
shown earlier in Table 2 of Ref. \cite{Nab09}). The effect of
lowering of centroid and increased total strength causes a
considerable enhancement in calculated  electron capture rate from
this first excited state. At a temperature of 3 GK, the calculated
value of electron capture rate from ground state is 0.35 s$^{-1}$
whereas the product of electron capture rate and occupation
probability of the state gives a total value of 0.33 s$^{-1}$. For
the same temperature, the calculated value of electron capture rate
from the first excited state is 5.75 s$^{-1}$ and the product of
electron capture rate and occupation probability for the first
excited state is 0.28 s$^{-1}$. This means that alone the first
excited state is contributing an additional 85$\%$ of the ground
state capture rate to the cumulated capture rate. The total electron
capture rate from all parent excited states is 0.71 s$^{-1}$ and the
ratio of ground state capture rate to total capture rate is thus
0.47. Taking the case of $^{58}$Ni (as a sample case for even-even
isotope, at the same temperature of 3 GK), the pn-QRPA calculated
ground state electron capture rate is 0.291 s$^{-1}$, the
contribution from first excited state (at 1.45 MeV) is 0.00133
s$^{-1}$. The total electron capture rate is 0.293 s$^{-1}$ and the
ratio of ground to total capture rate comes out to be 0.995.

At still higher stellar temperature of T$_{9} = 30 $K, excited state
rates contribute heavily for all isotopes of nickel (as can be seen
from Table~\ref{ta3}). As mentioned earlier all excited state GT
strength distributions were calculated in a microscopic fashion
within the framework of the pn-QRPA model and were significantly
different from their ground-state counterparts. The details of the
microscopic calculation of ground and excited state GT strength
distributions using the pn-QRPA model and phase space calculations
can be seen from Ref. \cite{Nab99}.

\begin{figure}[htbp]
\caption{(Color online) Half-lives for nickel isotopes as a function
of stellar temperature calculated at a density of $ \rho = 10^{8.5}$
gcm$^{-3}$.} \label{fig6}
\begin{center}
\begin{tabular}{c}
\includegraphics[width=0.6\textwidth]{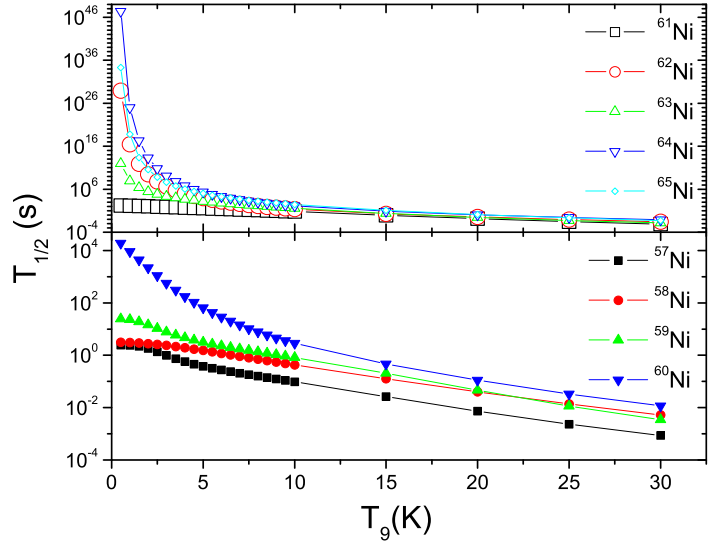}
\end{tabular}
\end{center}
\end{figure}

Figure~\ref{fig6} shows the calculated half-lives for nickel
isotopes as a function of stellar temperature at a selected density
of $\rho = 10^{8.5}$ gcm$^{-3}$. The total half-lives include
contribution from both electron capture and positron decay rates.
The lower panel shows the calculated half-lives for
$^{57,58,59,60}$Ni whereas the upper panel displays the calculated
half-lives for $^{61,62,63,64,65}$Ni. The calculated half-lives
decrease as stellar temperature increases as expected since the
electron capture rates increases substantially with increasing
temperatures (see Fig. \ref{fig4} and Fig. \ref{fig5}). The
half-life of $^{64}$Ni decreases by roughly 50 orders of magnitude
as the stellar temperature increases. This is primarily due to an
increase in the electron capture rates on $^{64}$Ni of similar
magnitude as stellar temperature increases.

In the next section I discuss how the pn-QRPA calculation compares
with the large scale shell model calculation for the astrophysically
important odd-A isotopes of nickel. This comparison might be of
special interest for core-collapse simulators.

\section{Comparison with shell model}
Table~\ref{ta4} compares the pn-QRPA calculated electron capture
rate on $^{57}$Ni with the large scale shell model calculation
(LSSM) \cite{Lan00}. All rates are given in units of $s^{-1}$. The
pn-QRPA calculated electron capture rates on $^{57}$Ni are generally
bigger than those calculated by LSSM. During the silicon burning
phases of massive stars the pn-QRPA calculated electron capture
rates are bigger by more than a factor of 6. The pn-QRPA model
calculated a total strength of magnitude 9.98 with a centroid around
7.25 MeV in daughter $^{57}$Co. At a stellar density of $\rho =
10^{7}$ gcm$^{-3}$ and a temperature of T$_{9} = 5 $K (these
physical conditions apply roughly to the silicon burning phases of
massive stars), the ground state contributes only about 10$\%$ to
the total electron capture rate. The remaining contribution comes
from excited states. As mentioned earlier Brink's hypothesis was not
assumed in the current calculation and all excited state
contributions were calculated in a microscopic fashion. At higher
densities, $\rho \sim 10^{9} - 10^{10} $ gcm$^{-3}$, relevant to
presupernova phase, the comparison improves.

The pn-QRPA electron capture rates on $^{59}$Ni are also bigger by
up to a factor 7 at low densities, $\rho \sim 10^{7} - 10^{8} $
gcm$^{-3}$ (see Table~\ref{ta5}). Collapse simulators should again
note that the pn-QRPA calculated electron capture rates on $^{59}$Ni
are around an order of magnitude bigger compared to LSSM rates
during the silicon burning phases of massive stars. At higher
stellar densities the two calculations are in excellent comparison.

The two calculations are in good comparison for the case of electron
captures on $^{61}$Ni as can be seen from Table~\ref{ta6}. During
the presupernova phase (densities around $\rho = 10^{10} $
gcm$^{-3}$) the LSSM rates are roughly double the corresponding
pn-QRPA rates for all temperature range shown in Table~\ref{ta6}.

Comparing electron capture rates on $^{63}$Ni, one notes from
Table~\ref{ta7} that LSSM numbers are twice bigger at high
temperature, T$_{9} = 10 $K, and high density, $\rho = 10^{10} $
gcm$^{-3}$ regions. Otherwise the two calculations are in excellent
agreement. Finally one notes from Table~\ref{ta8} that LSSM rates
are roughly twice the pn-QRPA rates at all temperature-density
domain shown in the table. For the case of $^{65}$Ni, the pn-QRPA
model calculated a total strength of magnitude 0.69 with a centroid
around 1.84 MeV in daughter $^{65}$Co. It is to be noted that during
the presupernova phase of massive stars the LSSM electron capture
rates on $^{63,65}$Ni are roughly twice the pn-QRPA calculated
rates.

\section{Conclusions}
The pn-QRPA model with an excellent track record of calculating
terrestrial half-lives was used to calculate electron capture and
positron decay rates on astrophysically  important isotopes of
nickel in stellar matter. A judicious choice of Gamow-Teller
strength parameters and use of experimental deformation parameter
for the even-even isotopes of nickel lead to an improved calculation
of the GT strength distributions for nickel isotopes. The Ikeda sum
rule was satisfied in the calculation and reasonable agreement was
achieved with the measured GT$_{+}$ centroids for the even-even
nickel isotopes. The finite temperature GT strength distributions
for all nickel isotopes were also calculated.

The positron decay rates of nickel isotopes are orders of magnitude
smaller than competing electron capture rates (except for the case
of $^{57}$Ni) and can be safely neglected in collapse simulations.
The positron decay rates of $^{57}$Ni competes well with the
electron capture rates only during the silicon burning phases of
massive stars. It was further shown that for stellar temperatures,
T$_{9}> 3 $K, electron captures on odd-A isotopes of nickel,
$^{57,59,61,63}$Ni, have dominant contributions from parent excited
states. Excited states contribute effectively for all isotopes of
nickel during presupernova phase of massive stars and beyond.

The electron capture rates on astrophysically important odd-A
isotopes were also compared with the LSSM calculation. For the case
of $^{61,63}$Ni the two calculations are in excellent agreement
except at high densities ($\rho \sim 10^{10} $gcm$^{-3}$) where
pn-QRPA rates are roughly half the corresponding LSSM rates. The
LSSM electron capture rates on $^{65}$Ni is twice the pn-QRPA rates.
During silicon burning phases of massive stars, the pn-QRPA
calculated electron capture rates on $^{57,59}$Ni are around an
order magnitude bigger whereas during presupernova phases the
calculated capture rates on $^{63,65}$Ni are half of the LSSM rates.
Collapse simulators are urged to take note of these comparisons for
a fine tuning of the lepton-to-baryon factor (and its time rate)
which is one of the key factors controlling the dynamics of
core-collapse.

\vspace{0.1in} \textbf{Acknowledgments}

The author wishes to acknowledge the support of research grant
provided by the Higher Education Commission, Pakistan, through HEC
Project No. 20-1283.

\newpage
\onecolumn
\begin{table}
\caption{Calculation of the total $S_{\beta^{\pm}}$ strengths in
nickel isotopes. The cut-off energy in daughter nuclei is 12 MeV.
The second column gives the values of nuclear deformation used in
the calculation.} \label{ta1} \scriptsize\begin{tabular}{cccc}
Nucleus &
Deformation & $\Sigma S_{\beta^{+}}$  & $\Sigma S_{\beta^{-}}$ \\
\hline
$^{57}$Ni & 0.03558 & 9.98$\times 10^{0}$ &  1.30$\times 10^{+1}$ \\
$^{58}$Ni & 0.18260 & 7.82$\times 10^{0}$ &  1.38$\times 10^{+1}$ \\
$^{59}$Ni & 0.02250 & 5.83$\times 10^{0}$ &  1.48$\times 10^{+1}$ \\
$^{60}$Ni & 0.20700 & 5.85$\times 10^{0}$ &  1.78$\times 10^{+1}$ \\
$^{61}$Ni &-0.09403 & 3.55$\times 10^{0}$ &  1.85$\times 10^{+1}$ \\
$^{62}$Ni & 0.19780 & 3.60$\times 10^{0}$ &  2.15$\times 10^{+1}$ \\
$^{63}$Ni & -0.09203& 1.73$\times 10^{0}$ &  2.27$\times 10^{+1}$ \\
$^{64}$Ni & 0.17900 & 1.78$\times 10^{0}$ &  2.57$\times 10^{+1}$ \\
$^{65}$Ni & -0.08054& 6.88$\times 10^{-1}$ & 2.75$\times 10^{+1}$ \\
\end{tabular}
\end{table}

\begin{table}
\caption{The pn-QRPA calculated centroids for nickel isotopes. Third
and fourth columns give the corresponding values calculated by Ref.
\cite{Lan00} and Ref. \cite{Pru03}, respectively (values adapted
from Ref. \cite{Pru03}). The last column shows measured values. For
references see text.} \label{ta2} \scriptsize\begin{tabular}{ccccc}
Nucleus & E$(GT_{+})$ & E$(GT_{+})$ [LMP]  & E$(GT_{+})$ [PF] &
E$(GT_{+})$ [exp]\\ \hline
$^{57}$Ni &  7.25$\times 10^{0}$  & - & - &-\\
$^{58}$Ni &  3.57$\times 10^{0}$  & 3.75$\times 10^{0}$  & 3.65$\times 10^{0}$  & 3.6 $\pm$ 0.2\\
$^{59}$Ni &  5.63$\times 10^{0}$  & - & - & -\\
$^{60}$Ni &  3.09$\times 10^{0}$  & 3.40$\times 10^{0}$  & 2.70$\times 10^{0}$  & 2.4 $\pm$ 0.3\\
$^{61}$Ni &  4.93$\times 10^{0}$  & 4.70$\times 10^{0}$  & 4.70$\times 10^{0}$  &- \\
$^{62}$Ni &  2.13$\times 10^{0}$  & 2.10$\times 10^{0}$  & 1.80$\times 10^{0}$  & 1.3 $\pm$ 0.3\\
$^{63}$Ni &  3.73$\times 10^{0}$  & - & - &- \\
$^{64}$Ni &  8.00$\times 10^{-1}$  & 1.30$\times 10^{0}$  & 1.80$\times 10^{0}$  & 0.8 $\pm$ 0.3\\
$^{65}$Ni &  1.84$\times 10^{0}$  & -& - &-\\
\end{tabular}
\end{table}

\begin{table}
\caption{The ratios of the ground state capture rates to total
capture rates, $R_{ec}(G/T)$ for isotopes of nickel. The first
column gives the corresponding values of stellar temperature,
$T_{9}$ (in units of $10^{9}$ K). The ratios are calculated at
density $10^{8.5} g cm^{-3}$.} \label{ta3} \scriptsize
\begin{tabular}{c|c|c|c}
 & \emph{$\mathbf{^{57}Ni}$}  & \emph{$\mathbf{^{58}Ni}$} & \emph{$\mathbf{^{59}Ni}$} \\
$\mathbf{T_{9}}$&
$\mathbf{R_{ec}(G/T)}$ &
$\mathbf{R_{ec}(G/T)}$ &
$\mathbf{R_{ec}(G/T)}$\\\hline
0.5 & 1.00$\times 10^{0}$    & 1.00$\times 10^{0}$   & 9.99$\times 10^{-1}$   \\
2   & 8.05$\times 10^{-1}$   & 1.00$\times 10^{0}$   & 5.25$\times 10^{-1}$   \\
3   & 4.70$\times 10^{-1}$   & 9.95$\times 10^{-1}$   & 2.61$\times 10^{-1}$   \\
5   & 2.20$\times 10^{-1}$   & 9.41$\times 10^{-1}$   & 1.32$\times 10^{-1}$   \\
10  & 1.29$\times 10^{-1}$   & 6.16$\times 10^{-1}$   & 1.42$\times 10^{-1}$   \\
30  & 1.39$\times 10^{-2}$   & 5.68$\times 10^{-2}$   & 2.01$\times
10^{-2}$
\\\hline
 & \emph{$\mathbf{^{60}Ni}$}  & \emph{$\mathbf{^{61}Ni}$} & \emph{$\mathbf{^{62}Ni}$} \\
0.5 & 1.00$\times 10^{0}$   & 8.14$\times 10^{-1}$   & 4.37$\times 10^{-2}$   \\
2   & 1.00$\times 10^{0}$   & 3.86$\times 10^{-1}$   & 5.30$\times 10^{-1}$   \\
3   & 9.92$\times 10^{-1}$   & 2.68$\times 10^{-1}$   & 6.85$\times 10^{-1}$   \\
5   & 9.26$\times 10^{-1}$   & 1.64$\times 10^{-1}$   & 6.45$\times 10^{-1}$   \\
10  & 5.48$\times 10^{-1}$   & 1.50$\times 10^{-1}$   & 3.83$\times 10^{-1}$   \\
30  & 4.98$\times 10^{-2}$   & 1.58$\times 10^{-2}$   & 4.34$\times
10^{-2}$
\\\hline
 & \emph{$\mathbf{^{63}Ni}$}  & \emph{$\mathbf{^{64}Ni}$} & \emph{$\mathbf{^{65}Ni}$} \\
0.5 & 1.03$\times 10^{-1}$   & 7.27$\times 10^{-1}$   & 7.55$\times 10^{-1}$   \\
2   & 8.91$\times 10^{-2}$   & 7.16$\times 10^{-1}$   & 7.01$\times 10^{-1}$   \\
3   & 1.11$\times 10^{-1}$   & 7.25$\times 10^{-1}$   & 6.61$\times 10^{-1}$   \\
5   & 1.65$\times 10^{-1}$   & 6.93$\times 10^{-1}$   & 5.79$\times 10^{-1}$   \\
10  & 1.74$\times 10^{-1}$   & 4.59$\times 10^{-1}$   & 3.31$\times 10^{-1}$   \\
30  & 1.21$\times 10^{-2}$   & 4.66$\times 10^{-2}$   & 1.53$\times
10^{-2}$
\end{tabular}
\end{table}

\begin{table}
\scriptsize{ \caption{The pn-QRPA and LSSM \cite{Lan00} calculated
electron capture rates, in units of s$^{-1}$, on $^{57}$Ni for
temperature and density domain of astrophysical interest.}
\label{ta4}
\begin{tabular}{c|cccccccc}   & pn-QRPA & LSSM
&pn-QRPA & LSSM & pn-QRPA & LSSM & pn-QRPA & LSSM \\
  $T_{9}$       & ($10^{7}$gcm$^{-3}$) & ($10^{7}$gcm$^{-3}$) & ($10^{8}$gcm$^{-3}$) & ($10^{8}$gcm$^{-3}$)&
       ($10^{9}$gcm$^{-3}$) & ($10^{9}$gcm$^{-3}$)& ($10^{10}$gcm$^{-3}$)& ($10^{10}$gcm$^{-3}$)\\\hline
2 &    1.95$\times 10^{-3}$   & 8.49$\times 10^{-4}$   &  4.25$\times 10^{-2}$   & 3.03$\times 10^{-2}$   & 4.28$\times 10^{0}$   & 7.48$\times 10^{0}$   & 9.29$\times 10^{+2}$   & 8.99$\times 10^{+2}$  \\
3 &    7.18$\times 10^{-3}$   & 1.32$\times 10^{-3}$   &  1.15$\times 10^{-1}$   & 4.35$\times 10^{-2}$   & 5.98$\times 10^{0}$   & 7.82$\times 10^{0}$   & 9.91$\times 10^{+2}$   & 8.97$\times 10^{+2}$  \\
5 &    2.91$\times 10^{-2}$   & 4.59$\times 10^{-3}$   &  3.78$\times 10^{-1}$   & 9.89$\times 10^{-2}$   & 1.19$\times 10^{+1}$   & 8.95$\times 10^{0}$   & 1.21$\times 10^{+3}$   & 8.95$\times 10^{+2}$  \\
10&    4.61$\times 10^{-1}$   & 1.50$\times 10^{-1}$   &
1.86$\times 10^{0}$    & 6.41$\times 10^{-1}$   & 3.58$\times
10^{+1}$   & 1.65$\times 10^{+1}$   & 1.85$\times 10^{+3}$   &
9.89$\times 10^{+2}$  \\
\end{tabular}}
\end{table}

\begin{table}
\scriptsize{\caption{Same as Table \ref{ta4} but for $^{59}$Ni.}
\label{ta5}
\begin{tabular}{c|cccccccc}   & pn-QRPA & LSSM
&pn-QRPA & LSSM & pn-QRPA & LSSM & pn-QRPA & LSSM \\
  $T_{9}$       & ($10^{7}$gcm$^{-3}$) & ($10^{7}$gcm$^{-3}$) & ($10^{8}$gcm$^{-3}$) & ($10^{8}$gcm$^{-3}$)&
       ($10^{9}$gcm$^{-3}$) & ($10^{9}$gcm$^{-3}$)& ($10^{10}$gcm$^{-3}$)& ($10^{10}$gcm$^{-3}$)\\\hline
2 &    3.06$\times 10^{-4}$   & 8.39$\times 10^{-5}$   &  6.41$\times 10^{-3}$   & 2.51$\times 10^{-3}$   & 8.13$\times 10^{-1}$  & 1.06$\times 10^{0}$ & 4.09$\times 10^{+2}$ & 4.26$\times 10^{+2}$\\
3 &    8.73$\times 10^{-4}$   & 1.51$\times 10^{-4}$   &  1.50$\times 10^{-2}$   & 4.02$\times 10^{-3}$   & 1.04$\times 10^{0}$   & 1.21$\times 10^{0}$ & 3.77$\times 10^{+2}$ & 4.23$\times 10^{+2}$\\
5 &    3.24$\times 10^{-3}$   & 4.95$\times 10^{-4}$   &  4.43$\times 10^{-2}$   & 1.07$\times 10^{-2}$   & 1.75$\times 10^{0}$   & 1.70$\times 10^{0}$ & 3.45$\times 10^{+2}$ & 4.31$\times 10^{+2}$\\
10&    5.11$\times 10^{-2}$   & 3.42$\times 10^{-2}$   &  2.10$\times 10^{-1}$   & 1.50$\times 10^{-1}$   & 4.67$\times 10^{0}$   & 4.86$\times 10^{0}$ & 3.86$\times 10^{+2}$ & 4.99$\times 10^{+2}$\\
\end{tabular}}
\end{table}

\begin{table}
\scriptsize{\caption{Same as Table \ref{ta4} but for $^{61}$Ni.}
\label{ta6}
\begin{tabular}{c|cccccccc}   & pn-QRPA & LSSM
&pn-QRPA & LSSM & pn-QRPA & LSSM & pn-QRPA & LSSM \\
  $T_{9}$       & ($10^{7}$gcm$^{-3}$) & ($10^{7}$gcm$^{-3}$) & ($10^{8}$gcm$^{-3}$) & ($10^{8}$gcm$^{-3}$)&
       ($10^{9}$gcm$^{-3}$) & ($10^{9}$gcm$^{-3}$)& ($10^{10}$gcm$^{-3}$)& ($10^{10}$gcm$^{-3}$)\\\hline
2 &    5.65$\times 10^{-7}$ & 4.67$\times 10^{-7}$ &  2.65$\times 10^{-4}$ & 1.72$\times 10^{-4}$ & 9.93$\times 10^{-2}$ & 8.28$\times 10^{-2}$ & 7.94$\times 10^{+1}$ & 1.43$\times 10^{+2}$\\
3 &    7.00$\times 10^{-6}$ & 5.04$\times 10^{-6}$ &  6.14$\times 10^{-4}$ & 3.92$\times 10^{-4}$ & 1.24$\times 10^{-1}$ & 1.07$\times 10^{-1}$ & 8.11$\times 10^{+1}$ & 1.46$\times 10^{+2}$\\
5 &    9.53$\times 10^{-5}$ & 6.15$\times 10^{-5}$ &  2.25$\times 10^{-3}$ & 1.51$\times 10^{-3}$ & 1.81$\times 10^{-1}$ & 2.00$\times 10^{-1}$ & 8.17$\times 10^{+1}$ & 1.51$\times 10^{+2}$\\
10&    5.77$\times 10^{-3}$ & 6.84$\times 10^{-3}$ &  2.47$\times 10^{-2}$ & 2.99$\times 10^{-2}$ & 6.64$\times 10^{-1}$ & 9.93$\times 10^{-1}$ & 9.71$\times 10^{+1}$ & 1.79$\times 10^{+2}$\\
\end{tabular}}
\end{table}

\begin{table}
\scriptsize{\caption{Same as Table \ref{ta4} but for $^{63}$Ni.}
\label{ta7}
\begin{tabular}{c|cccccccc}   & pn-QRPA & LSSM
&pn-QRPA & LSSM & pn-QRPA & LSSM & pn-QRPA & LSSM \\
  $T_{9}$       & ($10^{7}$gcm$^{-3}$) & ($10^{7}$gcm$^{-3}$) & ($10^{8}$gcm$^{-3}$) & ($10^{8}$gcm$^{-3}$)&
       ($10^{9}$gcm$^{-3}$) & ($10^{9}$gcm$^{-3}$)& ($10^{10}$gcm$^{-3}$)& ($10^{10}$gcm$^{-3}$)\\\hline
2 &    2.90$\times 10^{-12}$ & 4.30$\times 10^{-12}$ &  4.44$\times 10^{-9}$ & 6.10$\times 10^{-9}$ & 4.81$\times 10^{-3}$ & 5.70$\times 10^{-3}$ & 2.30$\times 10^{+1}$ & 4.98$\times 10^{+1}$\\
3 &    2.72$\times 10^{-9}$ & 3.44$\times 10^{-9}$ &  4.39$\times 10^{-7}$ & 5.21$\times 10^{-7}$ & 7.03$\times 10^{-3}$ & 7.62$\times 10^{-3}$ & 2.34$\times 10^{+1}$ & 4.98$\times 10^{+1}$\\
5 &    1.10$\times 10^{-6}$ & 1.37$\times 10^{-6}$ &  3.29$\times 10^{-5}$ & 3.97$\times 10^{-5}$ & 1.51$\times 10^{-2}$ & 1.81$\times 10^{-2}$ & 2.58$\times 10^{+1}$ & 5.28$\times 10^{+1}$\\
10&    9.08$\times 10^{-4}$ & 1.59$\times 10^{-3}$ &  4.02$\times 10^{-3}$ & 7.00$\times 10^{-3}$ & 1.40$\times 10^{-1}$ & 2.50$\times 10^{-1}$ & 4.17$\times 10^{+1}$ & 7.48$\times 10^{+1}$\\
\end{tabular}}
\end{table}

\begin{table}
\scriptsize{\caption{Same as Table \ref{ta4} but for $^{65}$Ni.}
\label{ta8}
\begin{tabular}{c|cccccccc}   & pn-QRPA & LSSM
&pn-QRPA & LSSM & pn-QRPA & LSSM & pn-QRPA & LSSM \\
  $T_{9}$       & ($10^{7}$gcm$^{-3}$) & ($10^{7}$gcm$^{-3}$) & ($10^{8}$gcm$^{-3}$) & ($10^{8}$gcm$^{-3}$)&
       ($10^{9}$gcm$^{-3}$) & ($10^{9}$gcm$^{-3}$)& ($10^{10}$gcm$^{-3}$)& ($10^{10}$gcm$^{-3}$)\\\hline
2 &    1.87$\times 10^{-17}$ & 3.69$\times 10^{-17}$ &  2.94$\times 10^{-14}$ & 5.51$\times 10^{-14}$ & 2.47$\times 10^{-7}$ & 3.71$\times 10^{-7}$ & 7.05$\times 10^{0}$ & 1.19$\times 10^{+1}$\\
3 &    1.25$\times 10^{-12}$ & 2.55$\times 10^{-12}$ &  2.13$\times 10^{-10}$ & 4.14$\times 10^{-10}$ & 9.44$\times 10^{-6}$ & 1.46$\times 10^{-5}$ & 6.89$\times 10^{0}$ & 1.21$\times 10^{+1}$\\
5 &    1.41$\times 10^{-8}$ & 3.26$\times 10^{-8}$ &  4.46$\times 10^{-7}$ & 9.93$\times 10^{-7}$ & 3.19$\times 10^{-4}$ & 5.93$\times 10^{-4}$ & 7.05$\times 10^{0}$ & 1.35$\times 10^{+1}$\\
10&    1.12$\times 10^{-4}$ & 2.89$\times 10^{-4}$ &  5.04$\times 10^{-4}$ & 1.28$\times 10^{-3}$ & 2.00$\times 10^{-2}$ & 4.72$\times 10^{-2}$ & 1.07$\times 10^{+1}$ & 2.19$\times 10^{+1}$\\
\end{tabular}}
\end{table}

\newpage

\end{document}